# A Directed -Threshold Multi-Signature Scheme

**Sunder Lal** [*] **and Manoj Kumar** [**]

*\* Dept of Mathematics, IBS Khandari.Dr. B.R.A.University Agra.*
Sunder_lal2@rediffmail.com.in.
*\*\* Dept of Mathematics, HCST, Farah – Mathura,* [*U. P.*] *– 281122.*
Yamu_balyan@yahoo.co.in

**Abstract.** In this paper, we propose a **Directed Threshold Multi-Signature Scheme.** In this threshold signature scheme, any malicious set of signers cannot impersonate any other set of signers to forge the signatures. In case of forgery, it is possible to trace the signing set. This threshold signature scheme is applicable when the message is sensitive to the signature receiver; and the signatures are generated by the cooperation of a number of people from a given group of senders.

## 1. Introduction

Digital signature is a cryptographic tool to authenticate electronic communications. Digital signature scheme allows a user with a public key and a corresponding private key to sign a document in such a way that anyone can verify the signature on the document (using her/his public key), but no one can forge the signature on any other document. This self-authentication is required for some applications of digital signatures such as certification by some authority.

In most situations, the signer is generally a single person. However, in some cases the message is sent by one organization and requires the approval or consent of several people. In these cases, the signature generation is done by more than one consenting person. A common example of this policy is a large bank transaction, by one organization, which requires the signature of more than one partner. Such a policy could be implemented by having a separate digital signature for every required signer, but this solution increases the effort to verify the message linearly with the number of signer. To solve this problems, Multisignature schemes [10,11,13,17,18] and threshold signature schemes [5,6,7,9,12,20] are used where more than one signers share the responsibility of signing messages

Threshold signatures are closely related to the concept of threshold cryptography, first introduced by Desmedt [5, 6, 7]. In 1991, Desmedt and Frankel [6] proposed the first (t, n) threshold digital signature scheme based on the RSA system.

In (*t, n*) threshold signature scheme, any subgroup of *t* or more shareholders of the designated group can generate a valid group signature in such a way that the verifier can check the validity of the signature without identifying the identities of the signers. In threshold schemes, when any *t* or more shareholders act in collusion, they can impersonate any other set of shareholders to forge the signatures. In this case, the malicious set of signers does not have any responsibility for the signatures and it is impossible to trace the signers. Unfortunately, with threshold schemes proposed so far, this problem cannot be solved.

In multisignature schemes, the signers of a multisignature are identified in the beginning and the validity of the multisignature has to be verified with the help of identities of the signers. For multisignatures, it is indeed unnecessary to put a threshold value to restrict the number of signers. Consider the situation, where a group of anonymous members would have to generate a multisignature. The members of this group use pseudonyms as their identities in the public directory. What concerns the verifier most is that a message is signed by at least *t* members and they indeed come from that group. Nevertheless, the verifier has no way to verify whether a user is in fact a member of that group because of the anonymity of the membership. In this case, the multisignature schemes cannot solve this problems, however, the threshold signature schemes do.

On the other hands, there are so many situations, when the signed message is sensitive to the signature receiver. Signatures on medical records, tax information and most personal/business transactions are such situations. Signatures used in such situations are called directed signatures [1, 2, 3, 14, 15, 19, 23, 24]. In directed signature scheme, the signature receiver has full control over the signature verification process and can prove the validity of the signature to any third party, whenever necessary. Nobody can check the validity of signature without his cooperation.

Combining these ideas, we propose a digital signature scheme named as Directed - threshold multi - signature scheme. The proposed scheme is based on Shamir's threshold signature scheme [21] and Schnorr signature scheme [22]. These basic tools are briefly described in the next section. This paper is organized as follows: -

The section-2 presents some basic tools. Section-3 proposes a Directed - threshold multi - signature scheme. Section-4 discusses the security of the Scheme. An illustration is discussed in section-5. Remarks are in section-6.

## 2. Preliminaries

Throughout this paper, we will use the following system setting.

- A prime modulous $p$, where $2^{511} < p < 2^{512}$;
- A prime modulous $q$, where $2^{159} < q < 2^{160}$ and $q$ is a divisor of $p - 1$;
- A number $g$, where $g \equiv k^{(p-1)/q} \bmod p$, $k$ is random integer with $1 \le k \le p - 1$ such that g >1; (g is a generator of order $q$ in $Zp^*$).
- A collision free one-way hash function $h$ [25];

The parameters *p, q, g* and *h* are publicly known. Every user has two keys one private and one public. We assume that a user A chooses a random $x_A \in Zq$ and computes $y_A = g^{x_A} \mod p$. He keeps $x_A$ as his private key and publishes $y_A$ as his public key.

In **Schnorr's signature scheme** the signature of the user A on the message *m* is given by $(r_A, S_A)$, where,

$$r_A = h(g^{k_A} \mod p, m), \text{ and } S_A = k_A - x_A \cdot r_A \mod p.$$

The signature are verified by checking the equality

$$r_A = h(g^{S_A} y^{r_A} \mod p, m).$$

**Threshold secret sharing scheme** is a scheme to distribute a secret key K into *n* users in such a way that any *t* users can cooperate to reconstruct K but a collusion of *t* – 1 or less users reveal nothing about the secret. There are many realization of this scheme, we shall use Shamir's scheme. This scheme is based on Lagrange's interpolation in a field. To implement it, a polynomial *f* of degree *t* – 1 is randomly chosen in Zq such that *f* (0) = K. Each user *i* is given a public identity $u_i$ and a secret share $f(u_i)$. Now any subset of *t* shareholders out of *n* shareholders can reconstruct the secret K = $f(0)$, by pooling their shares and using

$$f(0) = \sum_{i=1}^{t} f(u_i) \prod_{j=1, j \neq i}^{t} \frac{-u_j}{u_i - u_j} \mod q$$

Here we assume that the authorized subset of *t* users consists of shareholders *i* for *i* =1,2,3…t.

### 3. Directed - threshold multi - signature scheme

The signer of the conventional digital signature schemes is usually a single person. But when the message *m* **(sensitive to the signature receiver R)** is transmitted by an organization *S* to a person *R* and the message requires the approval of more than one person then the responsibility of signing the messages needs to be shared by a subset $H_S$ of *t* or more signer from a designated group $G_S$ of *n* users belonging to the organization *S*.

From the point of view of the signing group, what the group concerns is the **tracability of the signing subgroup and no one other than the signature receiver can check the validity of the signatures.** On the other hand, from the verifier point of view, **whether the signature is indeed from that group and signed by at least *t* members (not the membership of the members in that group) is the most important**. The signature receiver *R* can verify the signature and that *R* can prove its validity to any third party C, whenever necessary. Nobody can check the validity of the signature without the help of *R*. Both the **( *t* , *n*) threshold directed signature schemes** and the **multisignature schemes** couldn't solve this problem independently.

In this paper, we are going to combine the idea of **(*t*, *n*) threshold signature schemes and multisignature schemes with directed signature scheme** and propose a threshold signature scheme, called the **Directed - threshold– multisignature scheme**, to solve the above problem.

For our construction, we assume that there is a **trusted share distribution center (SDC) that** determines the group secret keys, all shareholders's secret shares, and a **designated combiner *DC*** who takes the responsibility to collect and verify each partial signature and then produce a group signature. This scheme consists of the following steps: -

### 3.1. Group Secret Key and Secret Shares Generation for the organization *S*

(a). **SDC** selects the group public parameters *p, q, g* and a collision free one way hash function *h* [25]. SDC also selects a polynomial

$$f_S(x) = a_0 + a_1 x + \ldots a_{t-1} x^{t-1} \bmod q, \text{ with } a_0 = x_S = f_S(0).$$

(b). **SDC** computes the group public key, $y_S$, as,

$$y_S = g^{f_S(0)} \bmod p.$$

(c). **SDC** randomly selects $K \in Z_q$ and computes a public value

$$W = g^{-K} \bmod p.$$

(d). **SDC** randomly selects $K_i \in Z_q$ and computes

$$l_i = [K_i + f_S(u_{S_i})] \bmod q.$$

**Here $u_{S_i}$ is the public value associated with each user *i* in the group $G_S$.**

(e). SDC computes a public value $v_{S_i}$ for each member of the group $G_S$, as,

$$v_{S_i} = l_i \cdot y_{S_i}^K \bmod p.$$

**Here $y_{S_i}$ is the public value associated with each user *i* in the group $G_S$.**

(f). SDC sends $\{v_{S_i}, W\}$ to each user *i* in the group $G_S$ through a public channel.

(g). SDC also needs to computes public values, $m_i$ and $n_i$, as ,

$$m_i = g^{l_i} \bmod p \text{ and } n_i = g^{K_i} \bmod p.$$

### 3.2. Partial Signature generation by any *t* users and verification

If any *t* members of the organization *S* out of *n* members agree to sign a message *m* for a person *R*. *R* possesses a pair ($x_R, y_R$). Then the signature generation has the following steps.

(a) Each user $i \in H_S$ randomly selects $K_{i_1}$ and $K_{i_2} \in Z_q$ and computes

$$u_i = g^{-K_{i_2}} \bmod p, \quad v_i = g^{K_{i_1}} \bmod p \text{ and } w_i = g^{K_{i_1}} y_R^{K_{i_2}} \bmod p.$$

(b) Each user makes $u_i$, $w_i$ publicly and $v_i$ secretly available to each member $i \in H_S$. Once all $u_i$, $v_i$ and $w_i$ are available, each member $i \in H_S$ computes the product $U_S$, $V_S$, $W_S$ and a hash value $R_S$, as,

$$U_S = \prod_{i \in H_S} u_i \bmod q, \quad V_S = \prod_{i \in H_S} v_i \bmod q,$$

$$W_S = \prod_{i \in H_S} w_i \bmod q \text{ and } R_S = h(V_S, m) \bmod q.$$

(c) Each user $i \in H_S$ recovers his/her secret share $l_i$, as,

$$l_i = v_{S_i} W^{x_{S_i}} \bmod p.$$

(d) Each user $i \in H_S$ modifies his/her shadow, as,

$$MS_{S_i} = l_i \cdot \prod_{j=1, j \neq i}^{t} \frac{-u_{S_j}}{u_{S_i} - u_{S_j}} \bmod q.$$

(e). Each user $i \in H_S$ uses his/her modified shadow $MS_{S_i}$ and computes a value $s_i$, as,

$$s_i = K_{i_1} + MS_{S_i} \cdot R_S \bmod q.$$

(e) Each member $i \in H_S$ sends his partial signature to the **designated combiner** *DC*.

(f) The *DC* verifies the partial signature ($s_i$, $v_i$, $R_S$) by the following congruence,

$$g^{s_i} \stackrel{?}{\equiv} v_i \cdot m_i \left( \prod_{j \in H_S, j \neq i} \frac{-u_{S_j}}{u_{S_i} - u_{S_j}} \bmod q \right)^{R_S} \bmod p.$$

If this congruence holds, then the partial signature ($s_i$, $v_i$, $R_S$) for shareholder $i$ is valid.

### 3.3. Group Signature generation

(a). *DC* can computes the group signature $S_S$ by combining all the partial signature, as,

$$S_S = \sum_{i \in H_S} s_i \bmod q.$$

**Here we should note that there is no secret information associated with the *DC*.**

(b). *DC* sends $\{S_S, U_S, W_S, m\}$ to $R$ as signature of the group $S$ on the message $m$.

### 3.4. Signature verification by R

To verify the validity of the group signature $\{S_S, U_S, W_S, m\}$ **the verifier R needs his/her secret key** $x_R$. This sub-section consists of the following steps: -

(a) The verifier R computes a verification value E, as,

$$E = \prod_{i \in H} n_i \left( \prod_{j \in H_S, j \neq i} \frac{0 - u_{S_j}}{u_{S_i} - u_{S_j}} \right) \bmod q \bmod p.$$

(b). The verifier R can recovers the values $R_R$ and $R_S$, as

$$R_R = W_S \cdot U_S^{x_R} \bmod p \quad \text{and} \quad R_S = h(R_R, m).$$

(c). The verifier R uses the following congruence to check the validity of the signature

$$g^{S_S} \stackrel{?}{\equiv} R_R \cdot (E \cdot y_S)^{R_S} \bmod p.$$

If this congruence holds, then the group signature $\{S_S, U_S, W_S, m\}$ is valid signature of the organization S on the message m.

### 3.5. Proof of validity by R to any third party C

(a). R computes $\mu = U_S^{x_R} \bmod p$ and $R_R = \mu \cdot W_S \bmod p$.

(b). R sends $\{R_R, E, S_S, U_S, m, \mu\}$ to C.

(c). C recovers $R_S = h(R_R, m)$ and uses the following congruence to check the validity of the signature

$$g^{S_S} \stackrel{?}{\equiv} R_R \cdot (E \cdot y_S)^{R_S} \bmod p.$$

If this does not hold C stops the process; otherwise goes to the next step.

(d) In a zero knowledge fashion R proves to C that $\log_{U_S} \mu = \log_g y_R$ as follows:-

- C chooses random $u, v \in Z_p$ computes $w = (U_S)^u \cdot g^v \bmod p$ and sends w to R.
- R chooses random $\alpha \in Z_p$ computes $\beta = w \cdot g^{\alpha} \bmod p$, $\gamma = \beta^{x_R} \bmod p$ and sends $\beta, \gamma$ to C.
- C sends u, v to R, by which R can verify that $w = (U_S)^u \cdot g^v \bmod p$.
- R sends $\alpha$ to C, by which she can verify that

$$\beta = (U_S)^u \cdot g^{v+\alpha} \bmod p, \quad \text{and} \quad \gamma = (\mu)^u \, y_R^{v+\alpha} \bmod p.$$

### 4. Security discussions

In this sub-section, we discuss the security aspects of proposed scheme. Here we discuss several possible attacks. But show that, none of these can successfully break our system.

**(a). Can any one retrieve the secret keys** $x_S = f_S(0)$ **from the group public key** $y_S$ **?**

This is as difficult as solving discrete logarithm problem. No one can get the secret key $x_S$, since $f_S$ is a randomly and secretly selected polynomial. On the other hand, by using the public keys $y_S$ no one can get the secret keys $x_S$ because this is as difficult as solving discrete logarithm problem.

**(b). Can one retrieve the secret shares, $f_S(u_{S_i})$ $i \in G_S$, from the equation**

$$v_{S_i} = f_S(u_{S_i}) \cdot y_{S_i}^K \mod p \ ?$$

No because $f_S$ is a randomly and secretly selected polynomial and $K$ is also a randomly and secretly selected integer by the SDC.

**(c). Can one retrieve the secret shares, $f_S(u_{S_i})$ $i \in G_S$, from the equation**

$$f_S(u_{S_i}) = v_{S_i} W^{x_{S_i}} \mod p \ ?$$

Only the user $i$ can recovers his secret shares, $f_S(u_{S_i})$, because $f_S$ is a randomly and secretly selected polynomial and $x_{S_i}$ is secret key of the user $i \in G_S$.

**(d). Can one retrieve the modified shadow $MS_{S_i}$, integer $K_{i_1}$, $R_S$ and partial signature $s_i$, $i \in G_S$ from the equation**

$$s_i = K_{i_1} + MS_{S_i} \cdot R_S \mod q \ ?$$

It is computationally infeasible for a forger to collect the $MS_{S_i}$, integer $K_{i_1}$, $R_S$ and partial signature $s_i, i \in G_S$.

**(e). Can the designated combiner $DC$ retrieve the any partial information from the equation,**

$$S_S = \sum_{i \in H_S} s_i \mod q \ ?$$

Obviously, this is computationally infeasible for $DC$.

**(f). Can one impersonate a user $i \in H$ ?**

A forger may try to impersonate a user $i \in H_S$, by randomly selecting two integers $K_{i_1}$ and $K_{i_2} \in Z_q$ and broadcasting $u_i$, $v_i$ and $w_i$. But without knowing the secret shares, $f_S(u_{S_i})$ and $R_S$, it is difficult to generate a valid partial signature $s_i$ to satisfy the verification equations,

$$g^{s_i} \stackrel{?}{\equiv} v_i \cdot m_i \left( \prod_{j \in H_S, j \neq i} \frac{-u_{S_j}}{u_{S_i} - u_{S_j}} \mod q \right)^{R_S} \mod p.$$

**(g). Can one forge a signature {$S_S$, $U_S$, $W_S$, $m$} by the following equation,**

$$g^{S_S} \stackrel{?}{\equiv} R_R \cdot (E \cdot y_S)^{R_S} \bmod p \ ?$$

A forger may randomly selects an integer $R_R$ and then computes the hash value $R_S$ such that $R_S = h(R_R, m) \bmod q$.

Obviously, to compute the integer $S_S$ is equivalent to solving the discrete logarithm problem. On the other hand, the forger can randomly select $R_S$ and $S_S$ first, then try to determine a value $R_R$ that satisfy the equation

$$g^{S_S} \stackrel{?}{\equiv} R_R \cdot (E \cdot y_S)^{R_S} \bmod p.$$

However, according to the property of the hash function $h$, it is quite impossible. Thus, this attack will not be successful.

**(h). Can $t$ or more shareholders act in collusion to reconstruct the polynomial $f_S(x)$ ?**

According to the equation,

$$f_S(x) = \sum_{i=1}^{t} f(u_{S_i}) \prod_{j=1, j \neq i}^{t} \frac{x - u_{S_j}}{u_{S_i} - u_{S_j}} \bmod q,$$

the secret polynomial $f_S$ can be reconstructed with the knowledge of any $t$ secret shares, $f_S(u_{S_i})$ $i \in G_S$.

But in our proposed scheme, the secret shares $l_i$, $i \in G_S$, contains the integer $K_i$ which only the trusted SDC knows and has to be removed first before reconstructing the polynomial $f_S(x)$. A malicious shareholder $i$ may try to retrieve the integer $K_i$ from the public key $n_i$. However, the difficulty is same as solving the discrete logarithm problem. Thus, any $t$ or more shareholders cannot conspire to reconstruct the polynomial $f_S(x)$ by providing their own secret shares.

**So, if in an organization the shareholders are known to each other, even than they cannot reconstruct the polynomial $f_S(x)$.**

## 5. Illustration

The following illustration is supporting our scheme for practical implementation. Suppose $|G_S| = 7$, $|H_S| = 5$, $p = 47$, $q = 23$, $g = 25$.

### 5.1. Group Secret Key and Secret Shares Generation for the organization S

(a). **SDC** selects a polynomial $f_S(x) = 13 + 18 x^4 \bmod 23$, so $x_S = 13$ and $y_S = 16$.

(b). **SDC** randomly selects $K = 14$ and computes a public value $W = 2$.

(c). **SDC** computes the secret key, public key, secret share and public value for each member of the group $G_S$, as shown in the following table.

| VALUE USER | Secret $x_i$ | Public $y_i$ | Secret $K_i$ | Secret $f_S(u_{S_i})$ | Secret $l_i$ | Public $m_i$ | Public $n_i$ | Public $v_i$ |
|---|---|---|---|---|---|---|---|---|
| User–$S_1$ | 13 | 16 | 8 | 2 | 10 | 3 | 17 | 41 |
| User–$S_2$ | 18 | 4 | 22 | 21 | 20 | 9 | 32 | 29 |
| User–$S_3$ | 19 | 6 | 10 | 16 | 3 | 21 | 3 | 1 |
| User–$S_4$ | 20 | 9 | 17 | 3 | 20 | 9 | 34 | 19 |
| User–$S_5$ | 17 | 34 | 14 | 14 | 5 | 12 | 24 | 38 |
| User–$S_6$ | 22 | 32 | 16 | 14 | 7 | 27 | 7 | 14 |
| User–$S_7$ | 15 | 36 | 21 | 22 | 20 | 9 | 37 | 44 |

### 5.2 Signature generation by any *t* users

If any five users $S_2, S_4, S_5, S_6$, and $S_7$ out of seven members agree to sign a message *m*, for a person *R* possessing $x_R = 9$, $y_R = 2$ then the signature generation has the following steps.

(a). $S_2$ randomly selects $K_{2_1} = 18$, $K_{2_2} = 17$ and computes $u_2 = 18$, $v_2 = 4$ and $w_2 = 3$.

(b). $S_4$ randomly selects $K_{4_1} = 17$, $K_{4_2} = 19$ and computes $u_4 = 8$, $v_4 = 34$ and $w_4 = 8$.

(c). $S_5$ randomly selects $K_{5_1} = 14$, $K_{5_2} = 13$ and computes $u_5 = 3$, $v_5 = 24$ and $w_4 = 7$.

(d). $S_6$ randomly selects $K_{6_1} = 19$, $K_{6_2} = 21$ and computes $u_6 = 14$, $v_6 = 6$ and $w_6 = 25$.

(e). $S_7$ randomly selects $K_{7_1} = 16$, $K_{7_2} = 18$ and computes $u_7 = 12$, $v_7 = 7$ and $w_7 = 34$.

(f). Each user computes the product $U_S = 8$, $V_S = 36$, $W_S = 14$ and $R_S = 9$ (let).

(g). $S_2$ recovers his/her secret share $l_2 = 20$ and computes $MS_{S_2} = 10$, $s_2 = 16$.

(h). $S_4$ recovers his/her secret share $l_4 = 20$ and computes $MS_{S_4} = 16$, $s_4 = 0$.

(i). $S_5$ recovers his/her secret share $l_5 = 5$ and computes $MS_{S_5} = 6$, $s_5 = 22$.

(j). $S_6$ recovers his/her secret share $l_6 = 7$ and computes $MS_{S_6} = 5$, $s_6 = 18$.

(k). $S_7$ recovers his/her secret share $l_7 = 20$ and computes $MS_{S_7} = 5$, $s_7 = 15$.

### 5.3. Partial signature verification and Signature generation by *DC*

(a) *DC* verifies each partial signature. For example for user $S_2$,

$$s_2 = 16, v_2 = 4, m_2 = 9, R_S = 9, \prod_{j=4,5,6,7. i=2} \frac{-u_{S_j}}{u_{S_i} - u_{S_j}} \mod 23 = 12.$$

He checks $25^{16} \stackrel{?}{\equiv} 4 \cdot (9^{12})^9 \mod 47$. This holds.

Similarly, he checks other partial signatures.

(b). *DC* computes a group value $S_S = 2$ and sends $\{2, 14, 8, m\}$ as signature of the group *S* for the message *m*.

### 5.4. Signature verification by the person *R*

(a). The verifier *R* computes a verification value $E = 18$.

(b). The verifier *R* can recovers the values $R_R = 36$ and $R_S = 9$.

(c). The verifier *R* uses the following congruence to check the validity of the signature

$$25^2 \stackrel{?}{\equiv} 36 \cdot (18.16)^9 \mod 47.$$

This congruence holds, so the group signature $\{2, 14, 8, m\}$ is valid signature of the group *S* on the message *m* for the person *R*.

### 5.5. Proof of validity by *R* to any third party C

(a). *R* computes $\mu = 16$, $R_R = 36$ and sends $\{36, 18, 2, 8, m, 16\}$ to C.

(b). C recovers $R_S = 9$ and check the concurrence $25^2 \stackrel{?}{\equiv} 36 \cdot (18.16)^9$ for the validity of the signature. This holds; so, C goes to the next steps.

(c). *R* in a zero knowledge fashion proves to C that $\log_8 16 = \log_{25} 2$ as follows:-

- C chooses random $u = 9$, $v = 11$ and computes $w = 25$ and sends $w$ to *R*.
- *R* chooses random $\alpha = 37$ computes $\beta = 36$, $\gamma = 9$ and sends $\beta$, $\gamma$ to C.
- C sends $u, v$ to *R*, by which *R* can verify that $w = 25$.
- *R* sends $\alpha$ to C, by which she can verify that $\beta = 36$ and $\gamma = 9$.

## 6. Remarks

In this paper, we have proposed a **Directed –Threshold Multi - signature Scheme.** There is a **trusted SDC** that determine the group secret keys, all shareholders's secret shares, and a **designated combiner *DC*** who takes the responsibility to collect and verify each partial signature and then produce a group signature, **but no secret information is associated with the *DC*.**

In this threshold signature scheme, any malicious set of signers cannot impersonate any other set of signers to forge the signatures. In case of forgery, it is possible to trace the signing set. Any *t* or more shareholders acting in collusion cannot conspire to reconstruct the polynomial $f_S(x)$ by providing their own secret shares and hence they cannot recover the group secret key.

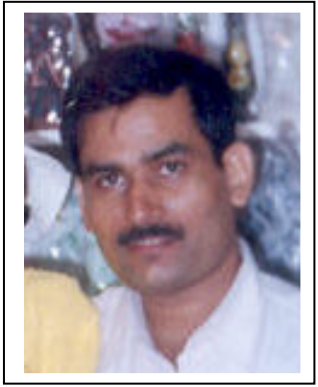

**Manoj Kumar** received the B.Sc. degree in mathematics from Meerut University Meerut, in 1993; the M. Sc. in Mathematics (Goldmedalist) from C.C.S.University Meerut, in 1995; the M.Phil. (Goldmedalist) in *Cryptography*, from Dr. B.R.A. University Agra, in 1996; submitted the Ph.D. thesis in *Cryptography*, in 2003. He also taught applied Mathematics at DAV College, Muzaffarnagar, India from Sep, 1999 to March, 2001; at S.D. College of Engineering & Technology, Muzaffarnagar, and U.P., India from March, 2001 to Nov, 2001; at Hindustan College of Science & Technology, Farah, Mathura, continue since Nov, 2001. He also qualified the *National Eligibility Test* (NET), conducted by *Council of Scientific and Industrial Research* (CSIR), New Delhi- India, in 2000. He is a member of Indian Mathematical Society, Indian Society of Mathematics and Mathematical Science, Ramanujan Mathematical society, and Cryptography Research Society of India. His current research interests include Cryptography, Numerical analysis, Pure and Applied Mathematics.